\documentclass{article}
\usepackage{arxiv}

\usepackage{amsmath,amsfonts,amssymb}
\usepackage{graphicx}
\usepackage{multirow}
\usepackage{authblk}
\usepackage[numbers,sort&compress]{natbib}
\usepackage[colorlinks=true,allcolors=blue,bookmarks=false]{hyperref}

\title{A Red Teaming Framework for Evaluating Robustness of AI-enabled Security Orchestration, Automation, and Response Systems}

\author[a]{Ayan Javeed Shaikh}
\author[b]{Nathaniel D.\ Bastian}
\author[a]{Ankit Shah\thanks{Corresponding author: \href{mailto:ankit@iu.edu}{ankit@iu.edu}}}
\affil[a]{Indiana University, Bloomington, IN, United States}
\affil[b]{United States Military Academy, West Point, NY, United States}

\date{}

\hypersetup{
  pdftitle={A Red Teaming Framework for Evaluating Robustness of AI-enabled Security Orchestration, Automation, and Response Systems},
  pdfauthor={Ayan Javeed Shaikh, Nathaniel D. Bastian, Ankit Shah},
  pdfkeywords={Red Teaming, Large Language Models, Reinforcement Learning, Cybersecurity, SOAR}
}

\begin{document}
\maketitle
\begin{abstract}
AI-enabled Security Orchestration, Automation, and Response (SOAR) systems increasingly employ autonomous agents for cyber defense, yet their resilience to adaptive adversaries is underexplored. We introduce an autonomous red teaming framework that integrates large language models (LLMs) with reinforcement learning (RL) to generate adaptive, multi-stage attack campaigns against autonomous defenders in enterprise networks. A hierarchical design combines an LLM-based planner for strategic intent with an RL controller for tactical execution, supported by reward shaping aligned with kill-chain progression. Evaluation in a high-fidelity enterprise simulation demonstrates the effectiveness of the proposed approach, while also showing that standalone LLM agents fail to sustain multi-stage attack campaigns and that domain-specific cybersecurity models achieve only limited levels of compromise, highlighting the necessity for hybrid LLM-RL approaches to red teaming.
\end{abstract}


\section{INTRODUCTION}
\label{sec:intro}
Cyber threats targeting enterprise networks continue to increase in both frequency and sophistication, with recent industry telemetry reporting on the order of approximately two thousand weekly attacks per organization globally in 2025, reflecting a sustained upward trend in adversarial activity \cite{checkpoint2026_report}. In response, enterprises are increasingly adopting Security Orchestration, Automation, and Response (SOAR) systems that leverage artificial intelligence (AI) to enable autonomous threat detection, incident response, and policy enforcement at machine speed \cite{shahriar2025agentic_security}. However, despite their growing deployment, these AI-enabled SOAR systems are not yet rigorously stress-tested against adaptive and strategic adversaries. Traditional penetration testing and rule-based red team scripts fail to capture the dynamic, multi-step nature of real-world attack campaigns, leaving a critical gap in evaluating the robustness of AI-enabled cyber defense systems~\cite{xu2025llm_cybersec_survey}. Autonomous red teaming, where an AI agent emulates an attacker following structured kill-chain behaviors such as reconnaissance, initial access, privilege escalation, lateral movement, and impact~\cite{mitre_attack}, offers a principled approach to closing this gap.

Recent advances in large language models (LLMs) have demonstrated strong capabilities for autonomous offensive cybersecurity tasks, including vulnerability discovery, threat reasoning, and multi-step attack planning \cite{xu2025llm_cybersec_survey, yang2025llm_aptds}. Domain-specific models such as Cisco's Foundation-Sec-8B \cite{foundation_sec} and DeepHat-7B, formerly WhiteRabbitNeo \cite{stark2024generative}, further improve structured reasoning in adversarial settings through security-focused fine-tuning. In parallel, reinforcement learning (RL) enables red teaming via interaction-driven policy optimization, learning sequential attack strategies that maximize long-horizon objectives in dynamic environments \cite{hore2025deep}. However, both paradigms exhibit fundamental limitations when applied independently to autonomous red teaming. LLM-based agents struggle with long-horizon state tracking, action consistency, and adaptation through environment feedback, while RL-based agents are constrained by sample inefficiency, limited interpretability, and poor generalization across heterogeneous cyber environments \cite{castro2025llm_acd}.

To address these limitations, we propose a hierarchical hybrid framework that integrates LLM-based strategic planning with RL-based tactical execution for red team operations. The LLM acts as a high-level planner that generates attack objectives, selects strategies, and modulates risk posture, while an RL controller executes low-level actions conditioned on environment state and planner directives. The LLM is kept frozen during training to preserve external knowledge, while RL optimizes execution policies through interaction. Optionally, a Reflexion mechanism \cite{shinn2023reflexion} enables iterative strategic refinement via episodic memory without parameter updates.

We evaluate our approach in the Cyber Operations Research Gym (CybORG) CAGE Challenge 4 environment \cite{baillie2020cyborg, cage4_paper}, a U.S. government–sponsored evaluation setting. CAGE~4 is a high-fidelity enterprise network simulation featuring partial observability, long-horizon decision-making, and adaptive defensive responses. The environment includes five coordinated blue team defenders powered by the Hierarchical Multi-Agent Reinforcement Learning (H-MARL) Expert policy,\cite{singh2025hierarchical} the strongest published autonomous defender for this setting, making it a rigorous benchmark for evaluating robustness under realistic adversarial conditions.

The key contributions of our work are as follows. We present a hierarchical LLM–RL architecture for autonomous red teaming that decouples strategic planning (LLM) from tactical execution (RL), enabling structured interaction between high-level intent generation and low-level environment-grounded control. We further introduce a kill-chain-aligned RL framework that incorporates structured reward shaping to guide long-horizon adversarial behavior consistent with MITRE ATT\&CK stages. In addition, we provide a comprehensive empirical study of LLMs across multiple families, including Qwen3, Llama, reasoning specialists such as DeepSeek-R1, and domain-specialized cybersecurity models such as Foundation-Sec and DeepHat, and across a wide parameter range (0.6B to 70B), revealing systematic limitations of standalone LLM agents in sustaining long-horizon cyber operations against adaptive autonomous defenders. Finally, we conduct an extensive evaluation in the CAGE~4 environment demonstrating the effectiveness of the proposed hybrid architecture under partial observability and multi-agent defender dynamics.

The remainder of this paper is organized as follows. Section~2 reviews related work in autonomous cyber operations, LLMs for cybersecurity, and hybrid LLM--RL architectures. Section~3 presents the proposed red teaming framework. Section~4 describes the experimental setup. Section~5 reports the results and compares the proposed approach with standalone LLM and RL baselines. Section~6 concludes with a discussion of limitations and directions for future work.

\section{LITERATURE REVIEW}
\label{sec:background}
This section reviews three bodies of literature that converge on the problem of autonomous red teaming: simulation environments for cyber operations, LLM applications in cybersecurity, and reinforcement learning for cyber agents. We then identify the gap that motivates our hierarchical architecture.

\subsection{Cyber Operations Simulation Environments}
\label{sec:bg_envs}
Evaluating autonomous cyber agents requires simulation environments that model realistic network topologies, multi-agent interaction, and partial observability. Several platforms have been developed to address this need. Microsoft's CyberBattleSim\cite{cyberbattlesim2021} provides a high-level abstraction of enterprise networks focused on post-breach lateral movement, using an OpenAI Gym interface to train RL agents; however, its abstract nature limits the fidelity of defensive responses. FARLAND\cite{molina2021farland} (Framework for Advanced Reinforcement Learning for Autonomous Network Defense), a MITRE--NSA collaboration, supports progressive complexity scaling and software-defined network reconfiguration for blue agent training, but focuses exclusively on defensive operations. CybORG\cite{baillie2020cyborg} introduced a dual-mode research gym, combining low-fidelity simulation for rapid training with high-fidelity emulation on real virtual machines supporting both red and blue team agents through an OpenAI Gym-compatible interface. Early experiments with Deep Double Q-Network (DDQN) agents demonstrated that RL could learn effective policies for simplified capture-the-flag scenarios within CybORG's simulation mode, but these experiments were limited to small networks with a single subnet and few hosts.

The CAGE Challenge series, built on CybORG, has progressively increased scenario complexity from single-subnet defense (CAGE~1) through enterprise-scale multi-agent operations (CAGE~4)\cite{cage4_paper}. CAGE~4 represents the most demanding configuration in the series: a multi-subnet enterprise network with coordinated blue team defenders operating across 500-step episodes under partial observability. Each episode features active defensive responses including host restoration, network isolation, and deception deployment that require the red agent to adapt its strategy in real time. CybORG has emerged as the standard benchmark for autonomous cyber agent research,\cite{castro2025llm_acd, shahriar2025agentic_security} and we adopt CAGE~4 as the primary evaluation environment for this work.

\subsection{Large Language Models for Cybersecurity}
\label{sec:bg_llm}
The application of LLMs to cybersecurity has expanded rapidly across multiple domains.
On the offensive side, LLM-driven penetration testing tools have demonstrated that language models can assist with vulnerability discovery and exploitation. PentestGPT\cite{deng2024pentestgpt} introduced a three-module architecture (Reasoning, Generation, and Parsing) that decomposes penetration testing into subtasks, achieving a 228.6\% task-completion improvement over GPT-3.5 baselines and earning the Distinguished Artifact Award at USENIX Security 2024. Another representative line of work includes WhiteRabbitNeo, a family of cybersecurity-focused generative models designed for offensive and defensive security tasks, which has evolved into the more recent DeepHat system, extending its capability toward uncensored red-team automation and improved inference efficiency in practical security operations \cite{stark2024generative}. HackSynth\cite{muzsai2024hacksynth} proposed a dual-module Planner-Summarizer architecture for autonomous CTF challenges, demonstrating that iterative command generation and feedback processing can solve challenges across diverse domains. Rigaki et al.\cite{rigaki2024stochastic} evaluated pre-trained LLMs as attacking agents in the NetSecGame environment using ReACT-style prompting, finding that zero-shot LLM agents achieve comparable or superior performance to RL agents trained for thousands of episodes in most scenarios. However, these systems rely on LLM reasoning alone without learned tactical policies, limiting their effectiveness in environments that require precise multi-step execution under active defense.

Domain-specialized security models have shown that targeted fine-tuning can close the performance gap with much larger general-purpose models. Foundation-Sec-8B\cite{foundation_sec} was trained on cybersecurity corpora including MITRE ATT\&CK mappings and threat intelligence datasets, achieving strong performance on security reasoning tasks at the 8B-parameter scale. Levi et al.\cite{levi2025cyberpal} introduced CyberPal~2.0, a family of cybersecurity-expert small language models (4B--20B parameters) that match or exceed GPT-4o on vulnerability-weakness correlation benchmarks, demonstrating that domain specialization at modest model sizes is viable for security applications. A notable finding from their work is that fine-tuning a base model yields 2.7$\times$ larger performance gains compared to fine-tuning an already instruction-tuned model, with direct implications for domain-specific LLM adaptation strategies. Yang et al.\cite{yang2025llm_aptds} proposed a dual-model collaborative architecture for Advanced Persistent Threat (APT) detection that iteratively refines threat assessments through multi-LLM reasoning aligned with MITRE ATT\&CK tactics, achieving F1 scores exceeding up to 99.20\% on standard APT detection datasets.

Despite these advances, LLMs face fundamental limitations as standalone autonomous cyber agents. Castro et al.\cite{castro2025llm_acd} evaluated LLM-based blue team agents in CAGE~4 and found that all-LLM teams achieved mean rewards approximately five times worse than all-RL teams ($-2{,}547$ vs.\ $-493$) while operating at roughly $1/104$th the decision-making speed. LLMs struggle with precise state tracking over hundreds of sequential decisions, exhibit action-repetition loops when operating in interactive environments, and cannot learn from trial-and-error interaction within a specific operational context. These findings motivate the search for architectures that preserve LLM reasoning capabilities while addressing their tactical execution limitations.

\subsection{Reinforcement Learning for Cyber Operations}
\label{sec:bg_rl}
Reinforcement learning has been applied to both offensive and defensive cyber operations, leveraging its ability to learn optimal sequential decision-making policies through environment interaction. The CAGE Challenge series has driven the development of increasingly sophisticated RL approaches. The Cybermonic KEEP submission\cite{king2025automated} employed graph convolutional networks (GCNs) with self-attention to process network topology as graph states, optimizing independent actor-critic agents per zone using PPO. Singh et al.\cite{singh2025hierarchical} proposed a hierarchical PPO architecture for CAGE~4 that decomposes the cyber defense task into specialized sub-policies for network investigation and host recovery, achieving top convergence speed and episodic return among CAGE~4 blue team submissions while enabling policy transfer across adversarial behavior shifts. We adopt their H-MARL Expert policy as the blue team defender in our evaluation, ensuring our red team agent is tested against the strongest published autonomous defense. 

However, RL for cyber operations faces persistent challenges. Pure RL agents require extensive training episodes to explore the large state-action space of realistic network environments, exhibit limited transferability across different network configurations, and produce opaque policies that resist human interpretation \cite{castro2025llm_acd}. The sparse reward structure of cyber operations where meaningful outcomes such as host compromise or service degradation occur only after long sequences of prerequisite actions exacerbates sample inefficiency and makes credit assignment difficult. Reward shaping aligned with domain knowledge (e.g., MITRE ATT\&CK kill chain progression) has been shown to improve RL training stability, but introduces the risk of reward hacking if not carefully constrained.\cite{xu2025llm_cybersec_survey}

On the offensive side, RL based red team agents remain significantly underexplored compared to defensive applications. The agentic security survey by Shahriar et al.\cite{shahriar2025agentic_security} found that only eight of over 160~surveyed papers employed RL or preference-based learning in security agent systems, and the vast majority of existing work focuses on defense rather than attack. This asymmetry is notable because robust evaluation of autonomous defenses requires equally capable autonomous adversaries precisely the capability our framework aims to provide.

\subsection{Hybrid Architectures and Research Gap}
\label{sec:bg_hybrid}
The complementary strengths and weaknesses of LLMs and RL have motivated growing interest in hybrid architectures across AI research. LLMs provide rich domain knowledge, strategic reasoning, and natural language explainability, but cannot learn from environment interaction. RL agents learn effective policies through trial and error, but require extensive training, produce opaque decisions, and struggle with long-horizon planning in the absence of domain priors.

Several works have begun exploring combinations of these paradigms in the cybersecurity domain. Castro et al.\cite{castro2025llm_acd} evaluated mixed teams of LLM and RL agents for blue team defense in CAGE~4, finding that heterogeneous teams, one LLM agent coordinating with four RL agents, could leverage both the reasoning capabilities of LLMs and the execution speed of RL. However, their approach combines agents at the \emph{team level}, assigning LLM and RL roles to different agents, rather than integrating both paradigms within a single agent's decision-making pipeline. Tholl et al.\cite{tholl2025large} investigated LLM integration with RL for autonomous cyber operations using the CAGE Challenge environment, finding that LLM-augmented reward shaping and action feedback can accelerate RL training and improve initial policy quality. Their work provides early evidence that LLMs can enhance RL-based cyber agents, though their integration operates at the reward and feedback level rather than through a structured planning hierarchy.

In the broader AI agent literature, hierarchical architectures that separate high-level planning from low-level control have shown success across robotics, game playing, and navigation tasks. The Reflexion framework\cite{shinn2023reflexion} demonstrated that LLM agents can improve their performance through verbal self-reflection and episodic memory without requiring weight updates, establishing the viability of learning-without-training paradigms for language model agents. The planner-executor pattern identified as the most common architecture in security agents\cite{shahriar2025agentic_security} typically uses LLMs for both planning and execution layers, missing the opportunity to leverage RL for adaptive tactical execution.

To the best of our knowledge, no prior work has proposed a hierarchical architecture that uses a frozen LLM as a strategic planner providing structured intent to a trainable RL controller for autonomous red team operations on AI-enabled cyber defense systems. While Singh et al.\cite{singh2025hierarchical} employ a hierarchical decomposition for RL-based defense, and Castro et al.\cite{castro2025llm_acd} combine LLMs and RL at the team level for defense, neither integrates LLM strategic reasoning with RL tactical learning within a single offensive agent. Existing LLM-based security agents lack the ability to improve through environment interaction; existing RL-based cyber agents lack strategic reasoning and domain knowledge priors. Our framework bridges this gap by decoupling strategic planning (LLM) from tactical execution (RL), allowing each component to operate in its area of strength: the LLM contributes knowledge of attack tactics and strategic reasoning about target selection, while the RL controller learns environment-specific tactical execution through thousands of training episodes guided by the planner's intent.

\section{Methodology}
\label{sec:method}

We propose a hierarchical framework for autonomous red teaming that integrates LLMs with RL to address long-horizon decision-making in adversarial environments. The framework decomposes the problem into two levels: (i) high-level strategic planning and (ii) low-level tactical execution, enabling complementary strengths of reasoning and learning to be combined within a unified agent.

\subsection{Problem Formulation}

We model autonomous red teaming as a partially observable Markov decision process (POMDP) defined by $(\mathcal{S}, \mathcal{A}, \mathcal{O}, \mathcal{T}, r, \gamma)$, where $s_t \in \mathcal{S}$ is the latent environment state, $o_t \in \mathcal{O}$ is the observation, $a_t \in \mathcal{A}$ is the action, and $r_t$ is the reward at time $t$. The agent interacts with the environment over long horizons, with the objective of maximizing cumulative reward while progressing through multi-stage adversarial objectives. Due to partial observability and delayed rewards, effective policies must reason over long temporal dependencies and maintain consistency across multi-step attack sequences.

\subsection{Hierarchical LLM--RL Architecture}
Our approach introduces a two-level hierarchy consisting of an LLM-based strategic planner and an RL-based tactical controller (see Figure~\ref{fig:architecture}).

\paragraph{Strategic Planner (LLM):}
The LLM operates at a coarse temporal resolution and produces a structured \emph{intent} conditioned on the current observation and auxiliary context. The intent encodes high-level decisions such as action type, MITRE ATT\&CK tactic, target selection, and risk posture. Formally, the planner defines:
\begin{equation}
    z_t = \pi_{\text{LLM}}(o_t, m_t),
\end{equation}
where $z_t$ is the structured intent and $m_t$ represents optional memory or contextual inputs. The LLM parameters remain fixed during training.

\paragraph{Tactical Controller (RL):}
The RL controller executes actions at every timestep, conditioned on both the current observation and the LLM-generated intent:
\begin{equation}
    a_t \sim \pi_{\theta}(a \mid o_t, z_t).
\end{equation}
This allows the controller to learn environment-grounded policies while following high-level strategic guidance.

To integrate both inputs, we encode observations into a latent representation $h_t$ and map the intent into a compatible embedding space. The combined representation is used by an actor-critic policy optimized via policy gradient methods. Notably, the planner operates at a lower frequency than the controller, generating intents every $h$ steps or when replanning is triggered. This reduces computational overhead while maintaining responsiveness to environment changes.

\subsection{State and Intent Representation}

Observations are transformed into two complementary representations:
\begin{itemize}
    \item A structured natural language summary used by the LLM planner.
    \item A numeric feature vector used by the RL controller.
\end{itemize}
The LLM outputs a structured intent object containing discrete and continuous attributes (e.g., action category, target, confidence), which is encoded into a fixed-dimensional embedding before being consumed by the RL controller.

\subsection{Hierarchical Reward Shaping}
To address sparse and delayed rewards, we introduce a novel reward function composed of multiple components:
\begin{equation}
    r_t = r_{\text{env}} + r_{\text{progress}} + r_{\text{constraint}} + r_{\text{alignment}},
\end{equation}
where $r_{\text{env}}$ is the environment reward that provides task-level feedback, $r_{\text{progress}}$ is the progress reward that encourages advancement through multi-stage objectives, $r_{\text{constraint}}$ represents constraint penalties that discourage invalid or out-of-order actions, and $r_{\text{alignment}}$ is the alignment reward that incentivizes consistency between RL actions and LLM intent. This decomposition is aimed towards improving credit assignment and stabilizes learning in long-horizon settings. Table~\ref{tab:reward_shaping} summarizes the reward components used to guide learning across different stages of the task.

\begin{table}[t]
\caption{Hierarchical reward shaping components used for training the RL controller.}
\label{tab:reward_shaping}
\centering
\begin{tabular}{l l l}
\hline
\textbf{Layer} & \textbf{Component} & \textbf{Reward} \\
\hline
L1: Environment & Negated defender reward & $r_{env} = -r_{\text{blue}}$ \\
\hline
\multirow{7}{*}{L2: Milestone}
& Discover new host & $+0.2$ \\
& Scan host & $+0.5$ \\
& Compromise (user access) & $+5.0$ \\
& Privilege escalation (root) & $+3.0$ \\
& Impact host & $+5.0$ \\
& Discover new subnet & $+1.0$ \\
& First compromise (defended subnet) & $+8.0$ \\
\hline
\multirow{3}{*}{L3: Constraint}
& Impact without root & $-0.1$ \\
& Exploit without scanning & $-0.05$ \\
& Escalate without user access & $-0.05$ \\
\hline
\multirow{2}{*}{L4: Intent Alignment}
& Match LLM intent & $+1.0$ \\
& Diverge from intent & $-0.3$ \\
\hline
\end{tabular}
\end{table}

\subsection{Policy Learning}
The RL controller is trained using an actor-critic method, Proximal Policy Optimization (PPO)\cite{schulman2017ppo}, to maximize expected return:
\begin{equation}
    J(\theta) = \mathbb{E}_{\pi_{\theta}} \left[ \sum_{t=0}^{T} \gamma^t r_t \right],
\end{equation}
where $\gamma$ is the discount factor and $T$ is the maximum number of timesteps. The controller receives both state and intent information, enabling it to learn when to follow or deviate from the planner based on environment feedback.

\subsection{Memory and Adaptation}
To enhance long-horizon reasoning, the planner can optionally maintain a bounded memory of prior interactions. In addition to short-term contextual inputs, our approach incorporates the Reflexion framework\cite{shinn2023reflexion}, wherein the planner generates a natural language self-reflection at the end of each episode. These reflections summarize successes, failures, and observed environmental patterns, capturing high-level strategic feedback such as ineffective actions, missed opportunities, and defender behaviors.

Reflections are stored in a bounded first-in, first-out (FIFO) buffer and incorporated into the planner’s input at the start of subsequent episodes. This provides cross-episode strategic context without modifying model parameters, enabling the frozen LLM to adapt its planning behavior through natural language feedback rather than gradient updates. This mechanism complements the RL controller’s weight-based learning. While policy optimization encodes tactical improvements through interaction data, the Reflexion framework enables rapid, explicit adjustments to high-level strategy, allowing the agent to improve both action selection and temporal decision-making across episodes.

\section{Experiments}
\label{sec:experiments}
\begin{figure*}[ht]
\centering
\includegraphics[width=\textwidth]{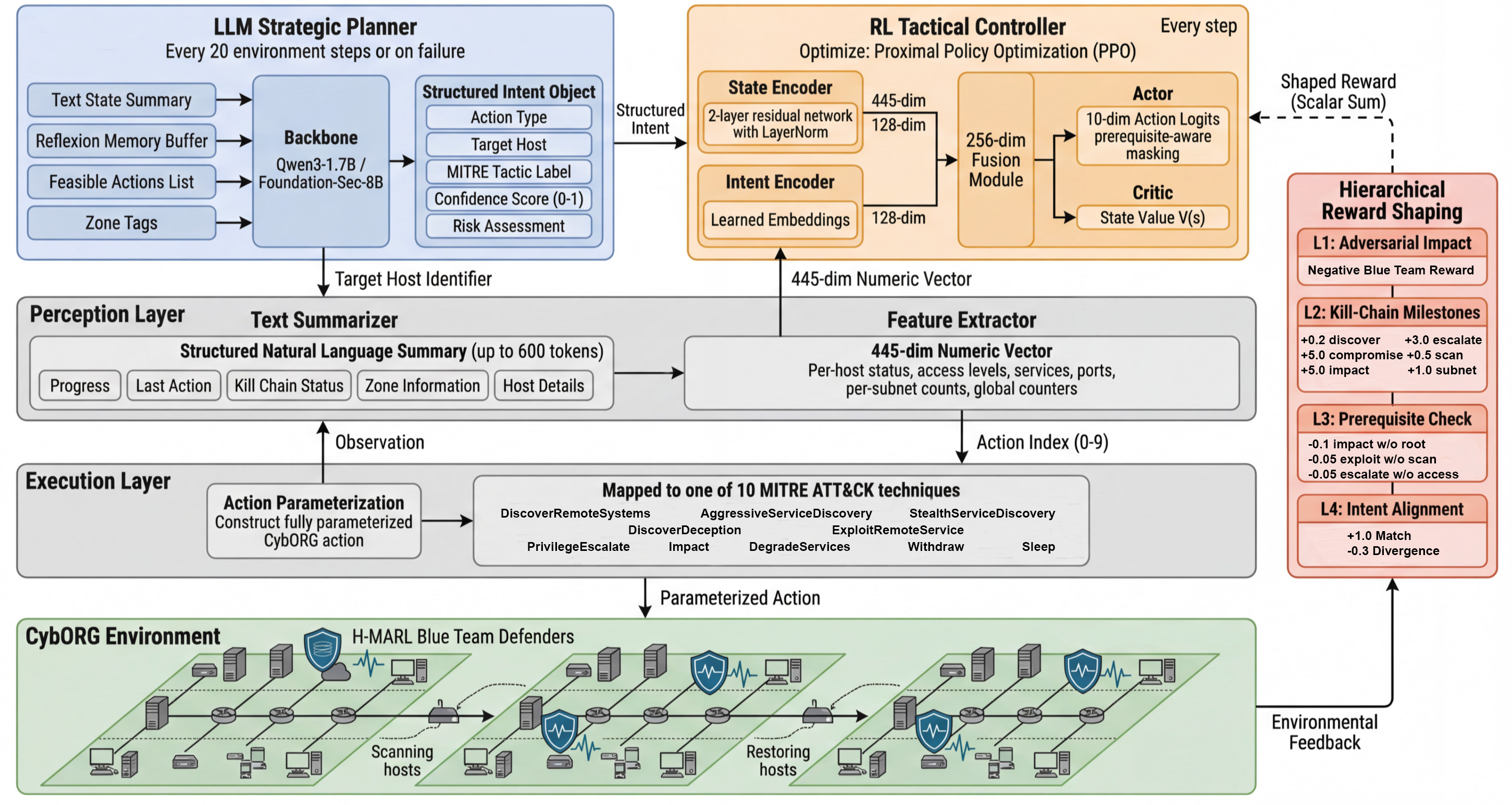}
\caption{Hierarchical LLM-RL red teaming framework architecture implemented in the CybORG CAGE~4 environment. The LLM planner provides strategic intent every 20 environment steps, while the RL controller selects tactical actions at every step. LLM weights remain fixed; only the RL controller is trained via PPO. The 4-layer reward shaping aligns policy optimization with kill-chain progression and strategic intent following.}
\label{fig:architecture}
\end{figure*}

We evaluate the proposed hierarchical LLM--RL framework in a high-fidelity cyber defense simulation to assess its effectiveness in generating long-horizon, multi-stage attack strategies against adaptive defenders. Figure~\ref{fig:architecture} illustrates the proposed red teaming architecture within a simulation-based testbed. Our experiments are designed to answer a key research question: whether integrating LLM-based strategic planning with RL-based execution improves autonomous red team performance against AI-enabled SOAR systems.

\subsection{Environment}
\begin{figure*}[ht]
\centering
\includegraphics[width=\textwidth]{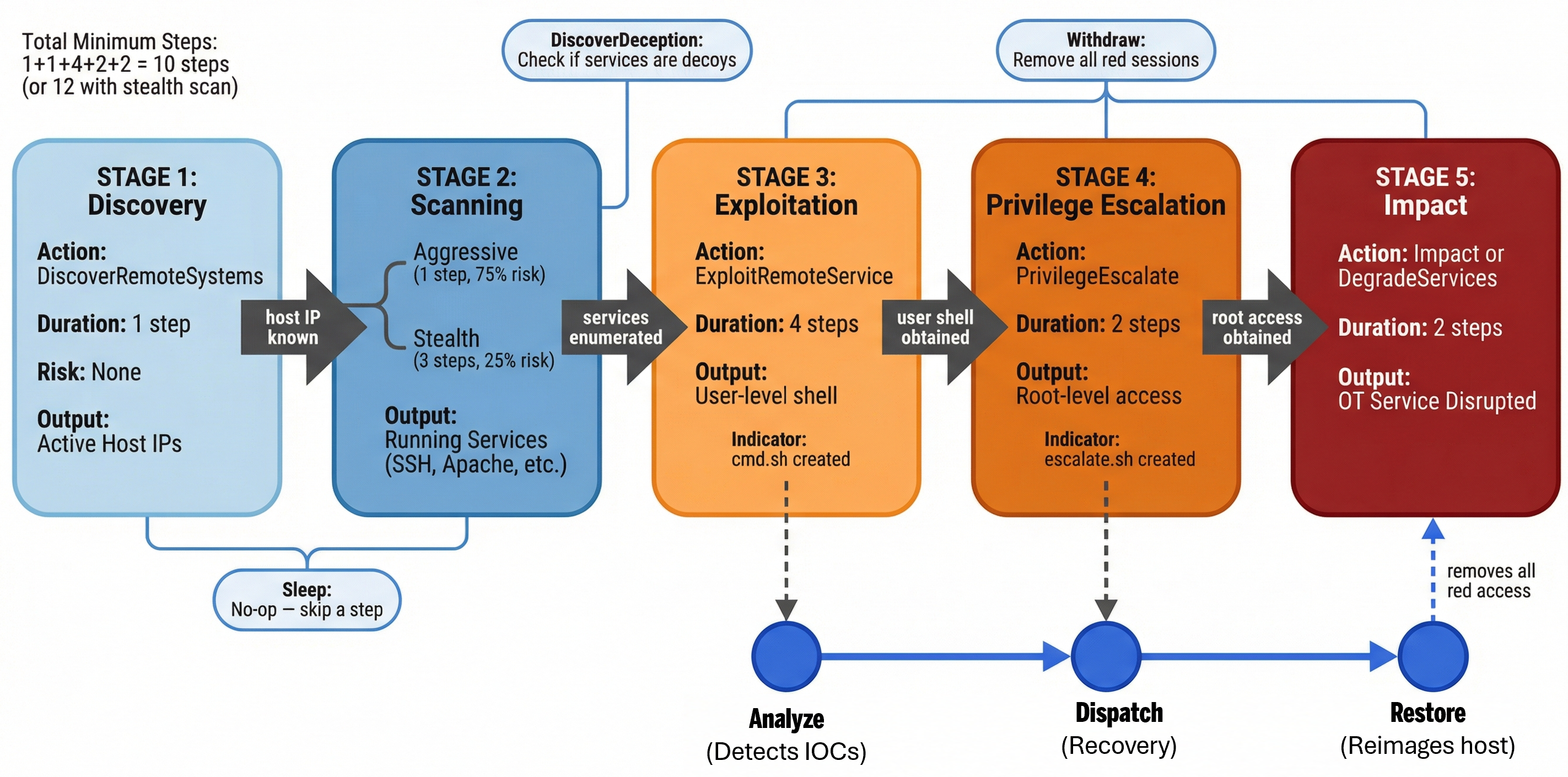}
\caption{MITRE ATT\&CK kill chain progression in CAGE~4. Each stage requires completion of the previous stage on the same host. The minimum path from discovery to impact requires 10 environment steps (12 with stealth scanning). Blue team defenders detect indicator files created during exploitation and escalation, triggering host restoration that removes all red team access.}
\label{fig:killchain}
\end{figure*}

We conduct all experiments in the CAGE~4 environment.\cite{baillie2020cyborg, cage4_paper} This environment simulates an enterprise network with multiple hosts, services, and subnets, and includes coordinated blue team defenders. The environment is partially observable and requires the red agent to perform multi-step operations such as reconnaissance, exploitation, privilege escalation, and impact over long horizons. Figure \ref{fig:killchain} shows a schematic of the kill chain progression in this environment. Each episode spans up to 500 timesteps. The blue team is controlled by an H-MARL policy \cite{singh2025hierarchical} representing a strong autonomous defender.

\subsection{Red Agent Configurations}
For our proposed hybrid configuration, the LLM planner operates at a planning horizon of $h=20$ steps and is also triggered upon action failure. At each timestep~$t$, the environment provides an observation~$o_t$ describing the current network state, including discovered hosts, available services, access levels, and active sessions. A perception layer (the CybORG adapter) converts this raw observation into a shared state representation consumed by both levels of the hierarchy: a text summary ($\leq$600 tokens) for the LLM planner, and a 445-dimensional feature vector for the RL controller.

The RL controller maps the shared state vector and the encoded LLM intent to a discrete action drawn from a 10-action space aligned with MITRE ATT\&CK tactics (see Table~\ref{tab:actions}). A state encoder compresses the observation into a 128-dimensional embedding via a two-layer residual network with layer normalization. When an LLM planner is present, an intent encoder maps the structured intent into a 128-dimensional vector, yielding a combined 256-dimensional input to the policy and value networks. A prerequisite-aware action mask constrains the RL controller to only select actions that have valid targets in the current state, preventing wasted actions while preserving autonomous decision-making (the RL chooses \emph{which} valid action to take, not a hand-coded rule).

The selected action index is translated into a CybORG-compatible action parameterized by the intent's target host and the current belief state in the execution layer. The environment returns the next observation, a reward signal, and a termination flag. A reward shaping module augments the sparse environment reward with kill-chain-aligned bonuses and penalties.

\begin{table}[ht]
\caption{Discrete action space aligned with MITRE ATT\&CK tactics. Actions 0--3 perform reconnaissance at different stealth levels, actions 4--5 advance the kill chain through exploitation and privilege escalation, actions 6--7 achieve operational impact, and actions 8--9 provide evasive maneuvers.}
\label{tab:actions}
\centering
\begin{tabular}{c l l}
\hline
\textbf{Index} & \textbf{Action} & \textbf{MITRE ATT\&CK Tactic} \\
\hline
0 & DiscoverRemoteSystems & Discovery \\
1 & AggressiveServiceDiscovery & Discovery \\
2 & StealthServiceDiscovery & Discovery \\
3 & DiscoverDeception & Discovery \\
4 & ExploitRemoteService & Initial Access \\
5 & PrivilegeEscalate & Privilege Escalation \\
6 & Impact & Impact \\
7 & DegradeServices & Impact \\
8 & Withdraw & Defense Evasion \\
9 & Sleep & Defense Evasion \\
\hline
\end{tabular}
\end{table}

We evaluate our approach by comparing with two other agent configurations to isolate the contribution of each component in our framework:
\begin{itemize}
    \item \textbf{LLM-only (zero-shot):} The LLM directly selects actions at each timestep without learning, establishing a baseline for purely reasoning-based agents. We evaluate 14 open-weight LLMs spanning four families and a wide range of parameter scales:
\begin{itemize}
    \item \textbf{General-purpose models:} Qwen3 (0.6B, 1.7B, 4B, 8B, 14B, 32B), a family of instruction-tuned models with native chain-of-thought reasoning capabilities; and Llama (Llama-3.1-8B-Instruct, Llama-3.3-70B-Instruct), widely adopted open-weight instruction-tuned models.
    \item \textbf{Reasoning-specialized models:} DeepSeek-R1 (14B) and QwQ (32B), optimized for extended chain-of-thought reasoning through distillation and specialized training.
    \item \textbf{Domain-specialized models:} Foundation-Sec-8B, DeepHat-7B, and RedSage-8B, fine-tuned on cybersecurity-specific corpora including MITRE ATT\&CK mappings, vulnerability analysis, and attack simulation data.
\end{itemize}
    \item \textbf{RL-only:} A PPO-based controller operates without LLM guidance, learning a policy solely from environment interaction.
\end{itemize}

\subsection{Training Details}

The RL controller is trained using PPO with Generalized Advantage Estimation (GAE) \cite{schulman2017ppo, schulman2016gae}. Training proceeds for up to 48 hours per run, with early stopping based on convergence of evaluation reward. In particular, training terminates when the evaluation reward improves by less than 0.5\% across five consecutive evaluation windows (checked every 1,000 episodes), subject to a minimum of 20,000 episodes to ensure sufficient exploration. We employ curriculum learning by gradually increasing environment difficulty and episode length. Table~\ref{tab:ppo_params} summarizes the hyperparameters, which were tuned through preliminary experiments in the CAGE~4 environment.

\begin{table}[ht]
\caption{PPO hyperparameters for RL controller training.}
\label{tab:ppo_params}
\centering
\begin{tabular}{l l}
\hline
\textbf{Parameter} & \textbf{Value} \\
\hline
Learning rate & $3 \times 10^{-4} \rightarrow 3 \times 10^{-5}$ (linear decay) \\
Clip ratio $\epsilon$ & 0.2 \\
Discount factor $\gamma$ & 0.99 \\
GAE parameter $\lambda$ & 0.95 \\
Mini-batch size & 64 \\
PPO epochs per rollout & 8 \\
Episodes per rollout & 10 \\
Max gradient norm & 0.5 \\
Entropy coefficient & 0.05 (Qwen3) / 0.10 (Foundation-Sec-8B) \\
Learning rate schedule & Cosine decay \\
\hline
\end{tabular}
\end{table}

\subsection{Evaluation Protocol and Metrics}
All experiments employ the H-MARL Expert policy~\cite{singh2025hierarchical} as the blue team defender. This policy comprises five coordinated AI agents that operate across network zones to perform investigation and recovery tasks. The expert master policy uses an Indicators of Compromise (IOC)-driven deterministic dispatch mechanism: upon detecting malicious files on a host, the corresponding agent transitions from investigation to recovery (i.e., host restoration).

We evaluate red team performance over 200 episodes using environment-native metrics that capture different stages of adversarial progress, which enable consistent comparison across different agent configurations and training regimes:
\begin{itemize}
    \item \textbf{Episode compromise rate (ECR)}: the fraction of episodes in which the agent achieves at least one host compromise through its own exploitation actions. This is our primary comparison metric, as it directly measures whether the agent can execute the attack kill chain.
    \item \textbf{Attack action count (AAC)}: the total number of exploit, escalation, and impact actions attempted across all episodes. 
    This behavioral metric distinguishes agents that actively generate attack campaigns from those that remain passive.
    \item \textbf{Average host discovery (AHD)}: for each episode, we record the maximum number of hosts the agent has discovered at any point during the episode; AHD is the mean of these per-episode maxima. This measures the agent's reconnaissance reach across the network.
\end{itemize}

\subsection{Experimental Design}
We design a set of controlled studies to isolate the contribution of key system components, focusing on three main axes: (i) language model choice, (ii) memory augmentation, and (iii) LLM–RL integration. We evaluate the following dimensions:
\begin{itemize}
    \item \textbf{LLM type:} General-purpose vs.\ domain-specialized models
    \item \textbf{Memory:} Reflexion-enabled vs.\ disabled variants
    \item \textbf{LLM–RL coupling:} RL with vs.\ without LLM-guided intent conditioning
\end{itemize}
These controlled variations allow us to quantify the impact of strategic reasoning (LLMs), episodic self-improvement (memory), and hybrid decision-making (LLM+RL).

Table~\ref{tab:phases} summarizes all evaluated configurations, grouped into three agent types: LLM-only, RL-only, and LLM+RL. Each configuration is evaluated under identical environmental conditions using stochastic action sampling against a fixed 5-agent H-MARL Expert blue team defender.

\begin{table*}[ht]
\caption{Overview of experimental configurations across LLM-only, RL-only, and LLM+RL agent classes.}
\label{tab:phases}
\centering
\small
\begin{tabular}{c l l c}
\hline
\textbf{Agent Type} & \textbf{Configuration} & \textbf{Purpose} & \textbf{Compute} \\
\hline

\multirow{4}{*}{LLM-only} 
 & Qwen3 (0.6B--32B, 6 models) & LLM scaling & 1--2$\times$H100 \\
 & Llama-3.1-8B, 3.3-70B & General & 1--4$\times$H100 \\
 & QwQ-32B, R1-14B & Reasoning & 1--2$\times$H100 \\
 & Cybersecurity models (Foundation-Sec, DeepHat, RedSage) & Domain & 1$\times$H100 \\
\hline

RL-only
 & PPO (RL only, no LLM) & RL baseline & 1$\times$H100 \\
\hline

\multirow{4}{*}{LLM+RL}
 & Qwen3-1.7B + PPO & General + RL & 2$\times$H100 \\
 & Qwen3-1.7B + PPO + Reflexion & + Reflection & 2$\times$H100 \\
 & Foundation-Sec-8B + PPO & Domain + RL & 2$\times$H100 \\
 & Foundation-Sec-8B + PPO + Reflexion & + Reflection & 2$\times$H100 \\
\hline

\end{tabular}
\end{table*}

\subsection{Implementation Details}
The LLM planner is served using \texttt{vLLM}~\cite{kwon2023vllm} or \texttt{HuggingFace Transformers}, selected based on model compatibility and serving efficiency. The RL controller is implemented as a standard actor–critic architecture with shared encoders that jointly process environment state and LLM-derived intent representations.

All experiments are executed on a GPU cluster with NVIDIA H100 (80 GB) GPUs. Each training run is allocated a fixed compute budget of up to 48 hours of wall-clock time to ensure consistent comparison across configurations. The framework is implemented in Python~3.11, using PyTorch~2.6 for RL training and vLLM~0.8.5 for high-throughput LLM inference.

For LLM-integrated configurations, the LLM is deployed as a local vLLM server on GPU~0, while RL training runs on GPU~1 within the same node, avoiding cross-node communication overhead. Both Foundation-Sec-8B-Reasoning and Qwen3-1.7B fit comfortably within a single H100 GPU under this setup. The LLM server operates asynchronously alongside training and is accessed through an OpenAI-compatible REST API over localhost. Ports are assigned dynamically to support multiple concurrent experiments without conflicts.

\section{Results}
\label{sec:results}
We first evaluate the proposed hierarchical LLM-RL framework and then compare it against standalone LLM and RL baselines.

\subsection{Evaluation of the Hierarchical LLM-RL Red Teaming Framework}
Table~\ref{tab:phase_d} summarizes the performance of the proposed framework across the top performing LLM backbones and memory configurations. The results demonstrate strong and consistent end-to-end attack capability across all configurations. In particular, the hybrid agents achieve near-perfect episode compromise rates, with Qwen3+RL attaining a 100\% success rate (200/200 episodes) and Foundation-Sec+RL achieving 94\%. These agents reliably progress through multiple stages of the attack sequence, including privilege escalation and impact, indicating effective long-horizon coordination between planning and execution.

\begin{table}[ht]
\caption{LLM+RL results (200 episodes).}
\label{tab:phase_d}
\centering
\small
\begin{tabular}{l c c c c}
\hline
\textbf{Configuration} & \textbf{AHD} & \textbf{ECR} & \textbf{AAC} \\
\hline
Qwen3 + RL            & 10.01 & \textbf{200 (100\%)} & 3,559 \\
Qwen3 + RL + Reflexion     & 10.01 & 199 (99.99\%) & 541 \\
Foundation-Sec + RL             & 9.86  & 188 (94\%) & 2,322 \\
Foundation-Sec + RL + Reflexion      & 9.97  & 182 (91\%) & 3,765 \\
\hline
\end{tabular}
\end{table}

The addition of the Reflexion framework maintains high success rates while altering behavioral efficiency, as reflected in differences in action counts. This suggests that reflection primarily influences strategic adaptation rather than raw success probability. Overall, the results show that the hierarchical decomposition enables stable and repeatable execution of multi-stage attack strategies in the presence of adaptive defenders.

\subsection{Comparison with Standalone LLM Agents}

Table~\ref{tab:llm_results} presents the performance of the top performing standalone LLM agents. In contrast to the hybrid approach, LLM-only agents exhibit limited end-to-end capability. Only six of 14 models achieve any compromise beyond the initial foothold, and the strongest model (QwQ-32B) succeeds in only 30\% of episodes. Moreover, successful behaviors are largely confined to early-stage reconnaissance and isolated exploitation attempts, without sustained progression through the attack chain.

\begin{table}[ht]
\caption{LLM-only results (200 episodes). Models with zero success are not shown.}
\label{tab:llm_results}
\centering
\small
\begin{tabular}{l l c c c}
\hline
\textbf{Model} & \textbf{Category} & \textbf{AHD} & \textbf{ECR} & \textbf{AAC} \\
\hline
QwQ-32B           & Reasoning-specialized  & 6.67  & \textbf{60 (30\%)} & 266 \\
DeepSeek-R1-14B   & Reasoning-specialized  & 10.17 & 26 (13\%) & 290 \\
Foundation-Sec-8B & Domain-specialized   & 10.13 & 13 (6.5\%) & 21,746 \\
DeepHat-7B        & Domain-specialized   & 10.01 & 8 (4\%) & 9,741 \\
Qwen3-0.6B        & General-purpose    & 3.74  & 7 (3.5\%) & 2,490 \\
Qwen3-1.7B        & General-purpose    & 10.09 & 2 (1\%) & 1,103 \\
\hline
\end{tabular}
\end{table}

\begin{figure}[ht]
\centering
\includegraphics[width=0.8\columnwidth]{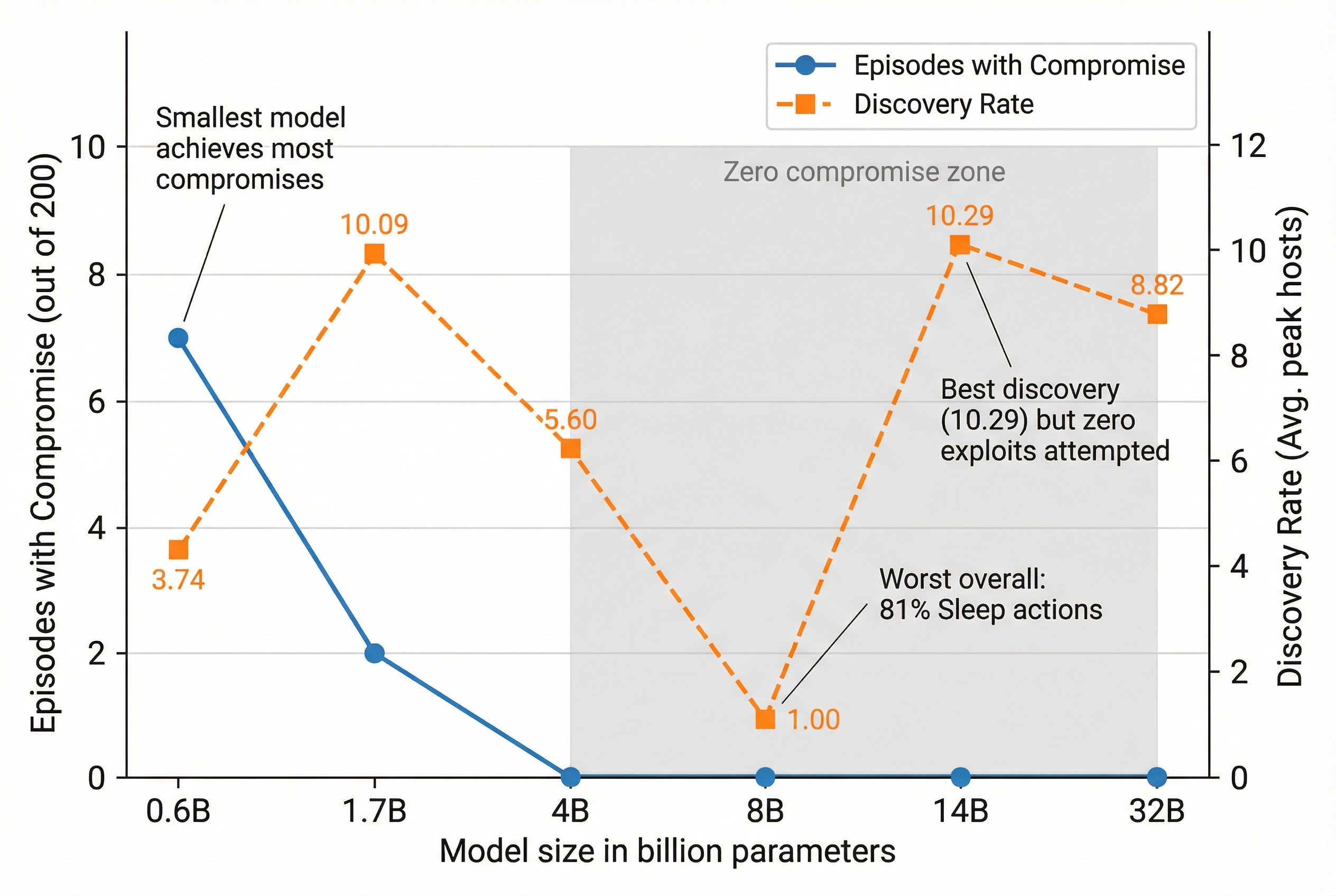}
\caption{Non-monotonic scaling in the Qwen3 model family. Left axis: episodes achieving compromise (out of 200). Right axis: average peak hosts discovered. Only the two smallest models achieve any compromise, while models at 4B+ never attempt exploitation.}
\label{fig:qwen3_scaling}
\end{figure}

Reasoning-specialized models outperform general-purpose and domain-specific models, achieving higher compromise rates. However, this advantage is confounded by model scale, as these models are also among the largest evaluated. Within the Qwen3 family, performance is non-monotonic with respect to model size: smaller models (e.g., 0.6B) achieve limited success, while larger variants fail to attempt exploitation altogether despite effective exploration  (see Figure \ref{fig:qwen3_scaling}).

Domain-specific models are not competitive despite generating substantially higher action volumes. For instance, Foundation-Sec-8B produces extremely large action counts in the LLM-only setting (21,746 actions) yet achieves only a 6.5\% ECR. This indicates that high domain-specific activity alone does not translate into meaningful attack progression, and that strategic reasoning plays a more critical role than domain specialization. These results indicate that while LLMs can provide useful strategic signals, they struggle with consistent multi-step execution and state tracking in interactive environments.

\subsection{Comparison with RL-only Agents}
The RL-only agent exhibits a degenerate policy collapse, selecting a single reconnaissance action (\textit{DiscoverRemoteSystems}) in nearly all timesteps. Across 200 evaluation episodes, it fails to achieve any successful compromise (0\% success rate). This behavior reflects a failure of reward-driven exploration to discover the structured sequence of actions required for multi-stage attacks, instead converging to a local optimum that maximizes immediate reward through repeated discovery actions.

We note that this outcome is influenced by the reward formulation used in our framework, which is designed to support the hierarchical LLM-RL setting. We do not perform additional reward tuning for the RL-only baseline in order to maintain a consistent and controlled comparison across agent configurations. As such, the observed collapse highlights the difficulty of learning effective multi-stage strategies under this reward structure without high-level guidance, rather than representing the best achievable performance of standalone RL under task-specific reward optimization.

\subsection{Key Insights}
These results highlight the complementary limitations of LLMs and RL when used in isolation. Standalone LLM agents provide high-level reasoning but fail to sustain coherent long-horizon behavior, while RL agents can optimize actions through interaction but fail to discover meaningful attack strategies due to sparse rewards and exploration challenges. Our proposed hybrid framework overcomes these limitations by combining strategic guidance from LLMs with environment-grounded learning in RL, resulting in substantial improvements in multi-stage attack success and robustness against adaptive defenses.

\section{CONCLUSION}
\label{sec:conclusion}
We present a hierarchical LLM-RL framework for evaluating the robustness of AI-enabled cyber defense systems in high-fidelity simulation. Across extensive experiments in the CybORG CAGE Challenge~4 environment, we demonstrate that autonomous red teaming requires the tight coupling of strategic reasoning and learned execution. Our principal findings from this study are as follows. First, standalone approaches are fundamentally insufficient. Across 14 LLMs, we observe limited end-to-end attack capability, with most models failing to sustain multi-step compromises. Similarly, RL alone collapses to a degenerate policy, unable to discover meaningful attack progression under sparse rewards. These results highlight the inability of either reasoning- or learning-only systems to execute full kill-chain behavior.

Second, integrating LLM planning with RL execution yields a multiplicative improvement. The hierarchical system achieves near-universal episode-level compromise and substantially higher root access rates than either component alone. The LLM provides high-level strategic intent over kill-chain progression, while the RL controller learns robust low-level execution, enabling sustained multi-step attack behavior under adaptive defense.

Third, we find that reliable structured planning is more critical than domain specialization. General-purpose models consistently outperform larger cybersecurity-specialized models, despite the latter producing significantly higher action volumes. This discrepancy is primarily driven by improved output reliability and fewer planning failures in general-purpose models, suggesting that structured controllability is a key requirement for LLMs in hierarchical decision systems.

Our study is constrained by several factors. First, evaluation is conducted in a single high-fidelity simulation environment (CAGE~4), and generalization to other network settings and real-world systems remains untested. Second, the framework uses a single red team agent, limiting analysis of coordination effects in multi-agent adversarial settings. Third, only a small subset of LLMs is integrated into the full hierarchical system due to computational constraints, leaving open the question of how stronger or larger models would affect performance. Finally, the environment introduces partial observability constraints that prevent full network topology awareness prior to host compromise, which may influence exploration dynamics.

Future directions include extending the framework to multi-agent red team coordination, improving topology inference through structured exploration strategies, and evaluating scalability across diverse environments and defender policies. We also plan to investigate larger and more capable LLM planners to better understand scaling behavior within hierarchical RL systems.

\section*{Acknowledgments}
This work was supported in part by the U.S. Military Academy (USMA) under Cooperative Agreement No. W911NF-25-2-0008 and the Defense Advanced Research Projects Agency (DARPA) under Cooperative Agreement No. HR0011-24-2-0004. The views and conclusions expressed in this paper are those of the authors and do not reflect the official policy or position of USMA, U.S. Army, U.S. Department of War, or U.S. Government.

\noindent Computational resources for this study were in part provided by Indiana University's Big Red~200 supercomputer.

\noindent Distribution statement: Approved for public release; distribution is unlimited.

\bibliographystyle{unsrtnat}
\bibliography{report}
\end{document}